\documentclass[fp]{jpsj3}
\usepackage{txfonts}
\usepackage{bm}
\usepackage{color}
\usepackage{graphicx}
\newcommand{\sigmav}{\bm \sigma}
\newcommand{\lambdav}{\bm \lambda}
\newcommand{\rv}{\bm r}
\newcommand{\nv}{\bm n}
\newcommand{\Jex}{J_{\rm ex}}
\newcommand{\average}[1]{\left\langle {#1} \right\rangle}
\newcommand{\MnFeGe}{Mn$_{1-x}$Fe$_x$Ge}
\newcommand{\FeCoGe}{Fe$_{1-x}$Co$_x$Ge}
\newcommand{\orb}{{l}}
\newcommand{\ra}{{\mu}}

\newcommand{\sa}{{\alpha}}
\renewcommand{\sb}{{\beta}}

\title{First-Principles Evaluation of the Dzyaloshinskii--Moriya Interaction}

\author{Takashi Koretsune$^{1,2}$\thanks{koretsune@cmpt.phys.tohoku.ac.jp}, Toru Kikuchi$^3$, and Ryotaro Arita$^4$}
\inst{$^1$Department of Physics, Tohoku University, Sendai 980-8578, Japan\\
$^2$JST PRESTO, 4-1-8 Honcho, Kawaguchi, Saitama 332-0012, Japan\\
$^3$Yukawa Institute for Theoretical Physics, Kyoto University, Kyoto 606-8502, Japan\\
$^4$RIKEN Center for Emergent Matter Science (CEMS), 2-1 Hirosawa, Wako, Saitama 351-0198, Japan \\
} 

\abst{
	We review recent developments of formulations to calculate the Dzyaloshinskii--Moriya (DM) interaction from first principles.
	In particular, we focus on three approaches.
	The first one evaluates the energy change due to the spin twisting by directly calculating the helical spin structure.
	The second one employs the spin gauge field technique to perform the derivative expansion with respect to the magnetic moment.
	This gives a clear picture that the DM interaction can be represented as the spin current in the equilibrium within the first order of the spin-orbit couplings.
	The third one is the perturbation expansion with respect to the exchange couplings and can be understood as the extension of the Ruderman--Kittel--Kasuya--Yosida (RKKY) interaction to the noncentrosymmetric spin-orbit systems.
	By calculating the DM interaction for the typical chiral ferromagnets \MnFeGe\ and \FeCoGe, we discuss how these approaches work in actual systems.
}

\begin{document}
\maketitle

\section{Introduction}
To control magnetic behavior for device applications, it is crucial to engineer magnetic interactions.
Usually, the interaction between spins has the symmetric form $\bm S_i \cdot \bm S_j$ and, as a result, the spins tend to be (anti)parallel.
On the other hand, in magnetic materials with broken inversion symmetry, a qualitatively different interaction, $\bm S_i \times \bm S_j$, called the Dzyaloshinskii--Moriya (DM) interaction, appears as a consequence of spin-orbit interactions\cite{dzyaloshinskii1958,moriya1960}. 
The DM interaction is antisymmetric for spin operators and favors twisted spin structures, which induces numerous interesting magnetic behaviors such as chiral soliton lattices in chiral helimagnets\cite{kishine2015}, skyrmion formation\cite{bogdanov1989,roszler2006,muhlbauer2009,yu2010,nagaosa2013}, and the enhancement of domain wall mobility\cite{thiaville2012,chen2013,ryu2013,emori2013,torrejon2014}.
In addition, the DM interaction relates magnetic properties and electric polarizations in multiferroic materials\cite{katsura2005,sergienko2006,tokura2014}.

In 1960, Moriya first proposed the microscopic derivation of the DM interaction at the first order of the spin-orbit coupling and discussed two different contributions\cite{moriya1960};
the first one is the extension of the superexchange mechanism to multiorbital spin-orbit systems, and the other is the combination of the direct exchange interaction and the spin-orbit coupling.
Although the physical pictures of these mechanisms are clear particularly for insulating systems, these formulations are not suitable for practical calculation.
For quantitative analysis, several techniques have been developed to calculate the DM interaction from first-principles calculations\cite{liechtenstein1987,igor1996,katsnelson2000,katsnelson2010,heide2008,ferriani2008,heide2009,freimuth2014,kikuchi2016,koretsune2015}.
Most of these approaches consider the energy difference by twisting the magnetic structures in various ways.
One of the most direct approaches is to calculate the energies of spirals with the finite vector $\bm q$ as $E(\bm q)$ and extract the $q$-linear term\cite{heide2008,ferriani2008,heide2009},
although this approach is sometimes time-consuming.
When the twisting angle is small, the energy change can be evaluated from the information of the uniform magnetic structures
by utilizing, for example, the magnetic force theorem\cite{liechtenstein1987,igor1996,katsnelson2000},
Berry phase\cite{freimuth2014}, or
spin gauge field transformation\cite{kikuchi2016}.
On the other hand, perturbation expansion with respect to the exchange couplings gives a different formulation to evaluate the DM interaction\cite{fert1980,imamura2004,kundu2015,wakatsuki2015,koretsune2015,shibuya2016}.

In this paper, we overview three approaches to evaluate the DM interaction from first-principles calculations,
that is, the methods using the energy of spirals, $E(\bm q)$, the spin current, and the off-diagonal spin susceptibility.
By applying these methods to chiral ferromagnets, namely, \MnFeGe\ and \FeCoGe\ systems, we discuss the relationship among these approaches,
how the band structures affect the DM interactions, and how each approach explains experimental results.

\section{Formulation to Compute the DM Interaction}
In this section, we describe three approaches to calculate the DM interaction.
Hereafter, we consider the low-energy effective Hamiltonian for the local direction of the magnetic continuum, $\bm n(\bm r)$, and neglect the charge degrees of freedom.
Then, the exchange interaction and the DM interaction are given as
\begin{align}
	H = \int d \bm r \;\sum_\mu\left[ J_\ra (\nabla_\ra \bm n)^2 + \sum_\alpha D^\sa_\ra (\nabla_\ra \bm n \times \bm n)^\sa\right],
	\label{eq:Hamiltonian_continuum}
\end{align}
where $J_\ra$ and $D_\ra^\sa$ denote the spin stiffness and the DM interaction, respectively.
Here, $\ra=x,y,z$ represents the relative direction of the two spins in the lattice representation, and $\sa=x,y,z$ represents the spin rotation axis.
This Hamiltonian can be derived from the local-spin Hamiltonian
\begin{align}
	H = \sum_{ij} \left[ - J_{ij} \bm S_i \cdot \bm S_j + \bm D_{ij} \cdot ( \bm S_i \times \bm S_j ) \right].
\end{align}
In fact, when we assume the isotropic system,
$J_\ra$ and $D_\ra^\sa$ can be written as
\begin{align}
	J &= \frac{1}{V}\sum_{j} \frac{1}{6}| \bm r_j |^2J_{0j},\\
	\bm D_\mu &= \frac{1}{V} \sum_{j} (-r_j)^\mu \bm D_{0j},
	\label{eq:JDcontinuum}
\end{align}
where $V$ is the volume of one site.
In this paper, we consider the total magnetic moment along the $z$-axis as a starting point and the DM interaction with $\sa=x,y$.


\subsection{Twisting magnetic structures}
Using the Hamiltonian in Eq.~\eqref{eq:Hamiltonian_continuum}, we can easily calculate the energy for a given $\bm n(\bm r)$.
For example, the exchange term always gives a non-negative value for $J_\ra > 0$ and the uniform magnetic structure, $\nabla_\ra \bm n = 0$, is most favorable.
On the other hand, the DM interaction term favors the twisted magnetic structure and its chirality depends on the sign of $D_\ra^\sa$.
In fact, when $J_\ra = J$ and $D_\ra^\sa = D \delta_{\ra \sa}$, a helical structure such as $\bm n = (\cos q z, \sin q z, 0)$ has $q$-dependent energy, $E(q) = J q^2 - D q$,
and the most stable wavevector is given as $q = D/2J$.
This means that once we can calculate the energy of the helical structure, $E(q)$, in actual systems, then we can evaluate the DM interaction as well as the spin stiffness in this effective Hamiltonian.
By changing the twisting axis and direction of $\bm n$, we can discuss each component of the DM interaction, $D_\ra^\sa$.

One way to compute $E(q)$ in the first-principles calculations is to employ the generalized Bloch theorem,
that is, apply different boundary conditions for up and down spins to simulate twisted magnetic structures\cite{heide2008,heide2009}.
Calculations of energies for actual twisted magnetic structures with supercells have also been performed\cite{yang2015}.

\subsection{Perturbation with respect to spin gauge field}
Here, we describe the spin current approach to the DM interaction using the method of the spin gauge field\cite{kikuchi2016}.
In Eq.~\eqref{eq:Hamiltonian_continuum}, electron degrees of freedom are integrated out and the Hamiltonian depends only on the magnetic moment $\bm n(\bm r)$.
To obtain this Hamiltonian from the microscopic model, we consider the following Hamiltonian in the field representation:
\begin{align}
	H &= \int d \bm r \; \sum_\orb c^\dagger_\orb \left[ -\frac{\hbar^2{\bm \nabla}^2}{2m} - \Jex^\orb {\bm n}\cdot \sigmav
  +\frac{i}{2}\sum_\ra \lambdav_\ra \cdot\sigmav \overleftrightarrow{\nabla}_\ra \right] c_\orb,
\label{eq:Hamiltonian}
\end{align}
where  $c^\dagger_\orb$ and $c_\orb$  are electron creation and annihilation operators for orbital $\orb$, respectively, 
$c^\dagger \overleftrightarrow{\nabla}_\ra c\equiv c^\dagger \nabla_\ra c- (\nabla_\ra c^\dagger) c$, and $\lambdav_\ra$ denotes the spin-orbit interaction.
The local direction of the magnetization $\bm n(\bm r)$, with $\bm n=(\sin\theta\cos\phi, \sin\theta\sin\phi, \cos\theta)$, is static, and $\Jex^\orb$ denotes the exchange constant.  
The form of $\lambdav_\ra$ is determined by the symmetries of the system; for example, $\lambda^\alpha_\mu \propto \delta^\alpha_\mu$ for B20 alloys such as MnSi and FeGe, while $\lambda^\alpha_\mu\propto \varepsilon_{\alpha\mu z}$ for Rashba systems with $z$ as its perpendicular direction.  
We consider here a simplified model with a quadratic dispersion and a spin--orbit interaction linear in the momentum but the extension to general cases is straightforward.

For this Hamiltonian, we consider the local spin gauge transformation so that the local magnetic moment $\nv(\rv)$ points to the $z$ direction.
For this purpose, we introduce a unitary transformation in spin space as
$c_\orb(\rv)=U(\rv)a_\orb(\rv)$, where $U$ is a $2\times2$ unitary matrix satisfying 
$U^\dagger(\nv\cdot\sigmav) U=\sigma^z$ \cite{TKS_PR08}. 
An explicit form of $U$ is given as $U=\bm m\cdot \bm \sigma$ with $\bm m\equiv (\sin\frac{\theta}{2}\cos\phi, \sin\frac{\theta}{2}\sin\phi, \cos\frac{\theta}{2})$. 
Geometrically, $U$ rotates the spin space by $\pi$ around the $\bm m$-axis.
Using this unitary transformation, derivatives of the electron field become covariant derivatives as  $\nabla_\ra c_\orb = U(\nabla_\ra + iA_{{\rm s},\ra})a_\orb$, where  
$A_{{\rm s},\ra} \equiv \sum_\sa A_{{\rm s},\ra}^\sa \frac{\sigma^\sa}{2} = -iU^\dagger \nabla_\ra U$ is an SU(2) gauge field, called a spin gauge field, given by $A_{{\rm s},\ra}^\sa=2(\bm m\times \nabla_\ra\bm m)^\sa$.
Then, the Hamiltonian for the electron in the rotated frame is given as $H=H_0+H_A$ with 
\begin{align}
	H_0\equiv & \int d \bm r \; \sum_{\orb} a^\dagger_{\orb} \left[- \frac{\hbar^2 {\bm \nabla}^2}{2m}  - \Jex^{\orb} \sigma^z + \frac{i}{2}\sum_\ra \tilde{\lambdav}_\ra \cdot \sigmav \overleftrightarrow{\nabla}_\ra   \right]a_{\orb},
\label{H0}\\
H_A \equiv & \int d\bm r \; \sum_{\orb}\left[\sum_{\ra \sa}\hat{\tilde{j}}_{{\rm s},\orb,\ra}^\alpha A^\sa_{{\rm s},\ra} +\frac{\hbar^2}{8m}\hat{n}_{{\rm el},\orb}(A_{{\rm s},\ra}^\sa)^2\right].
\label{HA}
\end{align}
Here, 
$\tilde{\lambda}^\sb_\ra\equiv \sum_\sa R_{\sa \sb}\lambda^\sa_\ra$ is the spin--orbit coupling constant rotated by the SO(3) matrix $R_{\sa \sb}\equiv 2m_\sa m_\sb - \delta^{\sa\sb}$ satisfying  $U^\dagger \sigma^\sa U=\sum_\sb R_{\sa \sb}\sigma^\sb$, and $\hat{n}_{{\rm el},\orb}\equiv a^\dagger_\orb a_\orb $.
The spin current density operator $\hat{\tilde{j}}_{{\rm s},\orb,\ra}^\sa$ in the rotated frame is given by $\hat{\tilde{j}}_{{\rm s},\orb,\ra}^\sa \equiv - \frac{i\hbar^2}{4m}a^\dagger_\orb \sigma^\sa \overleftrightarrow{\nabla}_\ra a_\orb - \frac{1}{2}\tilde{\lambda}^\alpha_\ra a_\orb^\dagger a_\orb$.

The interaction between electrons and the magnetization structure is originally given by the exchange interaction, $\Jex^\orb c^\dagger_\orb{\bm n}(\bm r)\cdot \sigmav c_\orb$ in Eq.~\eqref{eq:Hamiltonian}.  After the spin gauge transformation, 
this exchange interaction becomes a trivial one, $\Jex^\orb a^\dagger_\orb \sigma^z a_\orb$ in Eq.~\eqref{H0}.
The interaction between electrons and the magnetization structure is instead given by $H_A$ in Eq.~\eqref{HA}.  
We regard $H_0$ as the non-perturbative Hamiltonian and treat $H_A$ as a perturbation.
Since $A_{{\rm s},\ra}^\sa =2(\bm m\times \nabla_\ra \bm m)^\sa$ is proportional to the derivative of $\bm n(\bm r)$, the perturbative expansion by $H_A$ gives the derivative expansion with respect to the magnetization structure.  


Let us derive the effective Hamiltonian for magnetization, $H_{\rm eff}$, by integrating out the electron degrees of freedom in this rotated frame: 
\begin{equation}
\exp\left(-\frac{i}{\hbar}\int dt\; H_{\rm eff}\right) \equiv \int \mathcal D a^\dagger \mathcal D a \exp\left(\frac{i}{\hbar}(S_0-\int dt\; H_A)\right),
\label{Seff}
\end{equation}
where $S_0$ is the action corresponding to $H_0$, that is, $S_0\equiv \int dt[\int d\bm r\sum_\orb i\hbar a_\orb^\dagger \partial_t a_\orb - H_0]$.   We expand the right-hand side of Eq.~\eqref{Seff} by $H_A$, and obtain $H_{\rm eff}$ as 
\begin{align}
\int dt\; H_{\rm eff} =& i\hbar \ln Z_0 + \int dt \average{H_A} + \mathcal O((\partial \bm n)^2), 
\label{Seff2} \\
Z_0 \equiv& \int \mathcal D a^\dagger \mathcal D a \exp\left(\frac{i}{\hbar}S_0\right). 
\label{Z0}
\end{align}
Let us extract the DM interaction from $H_{\rm eff}$.  Since $\tilde{\lambda}^\sa_\ra(\bm r)=\sum_\sb R_{\sb \sa}(\bm r)\lambda^\sb_\ra$ in $H_0$ depends on the magnetization structure, not only $\average{H_A}$ but also $\ln Z_0$ contributes to the DM interaction.  However, the contribution from $\ln Z_0$ is only higher order in $\lambda$, while the contribution from $\average{H_A}$ contains first-order terms in $\lambda$.  Therefore, we can neglect the contribution from $\ln Z_0$ when $\lambda$ is sufficiently small.  
For $\average{H_A}$, since the DM interaction is the first-order derivative term in the effective Hamiltonian of the magnetization, it is sufficient to consider only the terms in $\average{H_A}$ proportional to $A_{\rm s}$.  Therefore, the DM interaction is included in   
\begin{align}
	H_{\rm eff} = & \int d\bm r \;   \sum_{\orb \ra \sa} \tilde{j}_{{\rm s},\orb,\ra}^\alpha A^\alpha_{{\rm s},\ra} ,
\label{Heff}
\end{align}
where $\tilde{j}_{{\rm s},\orb,\ra}^\alpha\equiv \average{ \hat{\tilde{j}}_{{\rm s},\orb,\ra}^\sa }$ is the expectation value of the spin current density in the rotated frame evaluated by $H_0$ in Eq.~\eqref{H0}. 
The spin current density $j_{{\rm s},\orb,\ra}^\sa$ in the laboratory frame is related as ${j}_{{\rm s},\orb,\ra}^\sa=\sum_\sb R_{\sa \sb}\tilde{j}_{{\rm s},\orb,\ra}^\sb$.
Then, by using the identity 
$\sum_\sb R_{\sa \sb}A_{{\rm s},\mu}^\sb = (\nabla_\ra \bm n\times \bm n)^\sa + n^\sa A_{{\rm s},\ra}^{z}
$,
the effective Hamiltonian reads 
\begin{align}
	H_{\rm eff} = & \int d\bm r \;  \left[\sum_{\ra \sa} D_\ra^\sa (\nabla_\ra \bm n\times \bm n)^\sa +  \sum_{\mu \orb} {j}_{{\rm s},\orb,\ra}^\parallel A_{{\rm s},\ra}^{z} \right],
\label{Heff2}
\end{align}
where 
${j}_{{\rm s},\orb,\ra}^\parallel \equiv \tilde{j}_{{\rm s},\orb,\ra}^z=\bm n\cdot\bm{j}_{{\rm s},\orb,\ra}$, and 
\begin{align}
        D_\ra^\sa \equiv
 \sum_{\orb} {j}_{{\rm s},\orb,\ra}^{\perp, \sa}
\label{eq:dm_spincurrent}
\end{align}
with $j_{{\rm s},\orb,\ra}^{\perp, \sa}\equiv {j}_{{\rm s},\orb,\ra}^\sa-n^\sa {j}_{{\rm s},\orb,\ra}^\parallel$.
Thus, the DM interaction is given by the expectation value of the spin current density of electrons.
More precisely, the transversely polarized component of the spin current $j_{{\rm s},\orb,\ra}^{\perp}$ contributes to the DM interaction.        
On the other hand, the longitudinally polarized component  $j_{{\rm s},\orb,\ra}^{\parallel}$ contributes to the spin-transfer torque term, $j_{{\rm s},\orb,\ra}^{\parallel}A_{{\rm s},\ra}^z=j_{{\rm s},\orb,\ra}^{\parallel}(1-\cos\theta)\nabla_\ra\phi$.  We have checked that, in the case of the simplified model in Eq.~\eqref{eq:Hamiltonian},  the contribution to the spin-transfer torque term from $j_{\rm s}^\parallel$ is cancelled by that from $\ln Z_0$ in Eq.~\eqref{Seff2} up to the $\lambda^3$-order.   This is expected since the spin-transfer torque will not arise spontaneously at equilibrium states.   

In the practical calculation, we consider a general form of the spin current density as
\begin{align}
	j_{{\rm s},\orb,\ra}^\sa = \sum_{\bm k}\frac{1}{4} \langle  c_{\bm k \orb}^\dagger ( v_\ra \sigma^\sa + \sigma^\sa v_\ra ) c_{\bm k \orb} \rangle,
	\label{J_DFT}
\end{align}
where the velocity operator is defined as $v_\ra = d H_{\bm k}/ d k_\ra$ with 
$H_{\bm k}=e^{-i\bm k\cdot \bm x}H e^{i\bm k\cdot \bm x}$.   
This general form of spin current actually corresponds to the DM interaction when we apply the spin gauge field technique to a generalized Hamiltonian instead of Eq.~\eqref{eq:Hamiltonian}.
Note that the DM interaction for the local-spin model derived from the tight-binding Hamiltonian\cite{katsnelson2010} corresponds to the discrete representation of Eqs.~\eqref{eq:dm_spincurrent} and \eqref{J_DFT}, which can be confirmed using Eq.~\eqref{eq:JDcontinuum}.

The spin current in Eq.~\eqref{eq:dm_spincurrent} is the expectation value at equilibrium states.  Spin currents at equilibrium states arise mainly by two mechanisms.
One is due to the magnetization structure, which induces a spin current of the form $j_{{\rm s},\ra}^\sa \propto (\bm n\times \nabla_\ra \bm n)^\sa$.
Such a magnetization-induced spin current is known to be relevant, for example, in multiferroic systems\cite{katsura2005}.
Since this spin current is proportional to $\nabla_\ra \bm n$, it does not contribute to the first-order derivative terms in Eq.~\eqref{Heff2}, but contributes to higher-order derivative terms in the effective Hamiltonian.
The other mechanism that induces a spin current at equilibrium states is the spin--orbit interaction with broken inversion symmetry, represented by the last term in Eq.~\eqref{eq:Hamiltonian}.  
This interaction tends to lock the relative angle of the spin and the momentum of electrons and yields a finite spin current.  The existence of such a spin--orbit-induced spin current has been noticed and discussed in the literature\cite{rashba2003,usaj2005,wang2006_2,sonin2007,sonin2007_prb,tokatly2008}. 
This spin--orbit-induced spin current arises even when $\nabla_\ra \bm n=0$ and  contributes to the first-order derivative terms in Eq.~\eqref{Heff2}.
Thus, to be precise, $j_{\rm s}^\perp$ in Eq.~\eqref{eq:dm_spincurrent} is generally not the total amount of spin current flowing in the system; when we expand a spin current in powers of $\nabla_\mu \bm n$, then $j_{\rm s}^\perp$ in Eq.~\eqref{eq:dm_spincurrent} corresponds to the non-derivative part.  In the practical calculation to evaluate, for example, the $D_\mu^x$ and $D_\mu^y$ components of the DM interaction, we set the magnetization direction $\bm n$ uniformly in the $z$-direction and calculate $j_{{\rm s},\mu}^x$ and $j_{{\rm s},\mu}^y$, respectively.

\begin{figure}
	\includegraphics[bb=0 0 1224 738, width=0.45\textwidth]{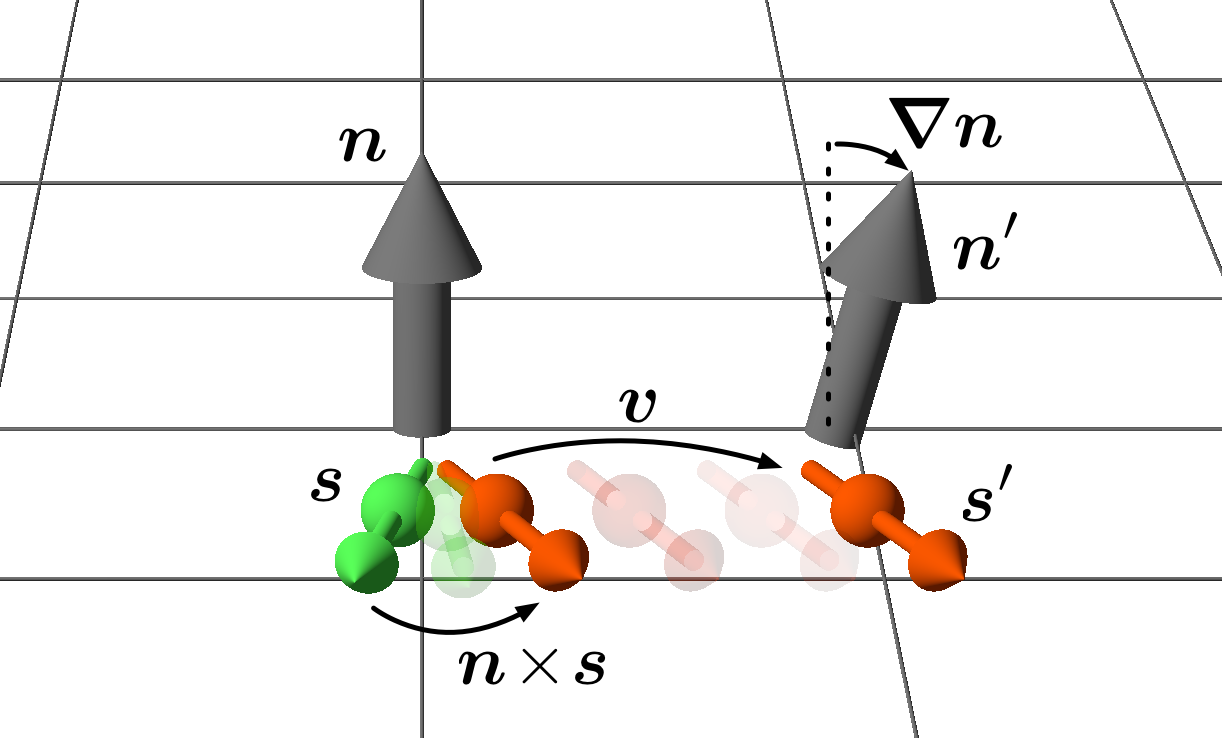}
	\caption{
		(Color online) Schematic picture showing the mechanism of the spin-current-induced Doppler shift.		
 A spin current $j_{{\rm s},\mu}^\alpha=s^\alpha v_\mu$ is flowing with spin polarization $\bm s$ and velocity $\bm v$.  Due to the torque from the localized spin $\bm n$, this spin current flows with its spin precessing.  Such a spin current with precession changes the spatial variation of the localized spins from $\nabla_\mu \bm n$ to $\nabla_\mu \bm n - \bm j_{{\rm s},\mu}\times \bm n$, which is the Doppler shift due to the spin current.    
 Adapted with permission.\cite{kikuchi2016} Copyright 2016, American Physical Society.
	}
\label{fig:Doppler}
\end{figure}

Our result, Eq.~\eqref{eq:dm_spincurrent}, clarifies that a spin current is a direct origin of the DM interaction.  Let us give a physical interpretation of this result.  Generally,  when an interaction is mediated by some medium, the interaction changes with its flow.  This phenomenon is known as the Doppler effect.  In the case of magnets, magnetic interactions are mediated by electron spin hopping among magnetic moments.  Therefore, when the electron spin flows as a spin current, the magnetic interaction changes and an additional interaction will emerge.   Since the spin current makes two adjacent magnetic moments inequivalent in the sense that one is located upstream and the other downstream,  the additional magnetic interaction is antisymmetric with respect to the exchange of the two adjacent magnetic moments.  Thus, an antisymmetric magnetic interaction, that is, the DM interaction, arises as the Doppler effect due to the spin current.  Let us see this Doppler effect more closely in Fig.~\ref{fig:Doppler}.  Consider two adjacent magnetic moments with directions $\bm n$ and $\bm n'=\bm n+ (\bm a\cdot \bm \nabla)\bm n$, where $\bm a$ is the vector connecting the two sites, and electrons hopping between them.  When an electron with spin $\bm s$ hops from $\bm n$ to $\bm n'$,  its spin precesses by $\epsilon (\bm s\times \bm n)$, with $\epsilon$ a small coefficient, due to the torque from the magnetic moments.  For the electron hopping with such a precession, the spatial variation of the magnetic moments will look like not $(\bm a\cdot \bm \nabla)\bm n$ but rather $(\bm a\cdot \bm \nabla)\bm n - \epsilon (\bm s\times \bm n)$.  When we express a spin current as $j_{{\rm s},\mu}^\alpha=s^\alpha v_\mu$ with $v_\mu$ the velocity of an electron, the discussion so far suggests that the derivative of the magnetization vector changes from $\nabla_\mu \bm n$ to $\mathfrak{D}_\mu\bm n\equiv\nabla_\mu\bm n-\eta(\bm j_{{\rm s},\mu}\times \bm n)$, with $\eta$ some coefficient, due to the presence of the spin current.  Accordingly,  the interaction energy, which was originally  $(\nabla_\mu \bm n)^2$ when there are no spin currents, changes into  
$(\mathfrak{D}_\mu\bm n)^2\cong (\nabla_\mu\bm n)^2+2\sum_\mu\eta\bm j_{{\rm s},\mu}\cdot(\nabla_\mu\bm n\times \bm n)$.  Thus, the DM interaction $\sum_\mu\bm D_\mu\cdot(\nabla_\mu\bm n\times \bm n)$ arises as the spin-current-induced Doppler shift of the exchange energy $(\nabla_\mu \bm n)^2$.  
Note that only the spin current component $j_{\rm s}^\perp$ perpendicular to $\bm n$ contributes to the Doppler shift $\mathfrak{D}_\mu\bm n\equiv\nabla_\mu\bm n-\eta(\bm j_{{\rm s},\mu}\times \bm n)$, which is consistent with the result of Eq.~\eqref{eq:dm_spincurrent}.  

Finally, let us mention that the spin current method can be applied to insulating systems.
In fact, the intrinsic spin current can be finite even in insulators.
The application to insulators such as Cu$_2$OSeO$_3$ will be discussed elsewhere.

\subsection{Perturbation with respect to exchange couplings}
Here, we consider Eq.~\eqref{eq:Hamiltonian} directly to derive the DM interaction.
We assume that $\Jex$ is small and consider the second-order perturbation to integrate out the electron degrees of freedom as in the derivation of the RKKY interaction.
Since there is a spin-orbit interaction, the effective Hamiltonian for $\bm n(\bm r)$ includes the DM interaction term $D_\ra^\sa (\nabla_\ra \bm n\times \bm n)^\sa$ with\cite{koretsune2015}
\begin{align}
	D_\ra^\sb = \frac{\Jex^2}{2} 
	\lim_{q \to 0} \frac{\partial \chi_0^{\sa \gamma}(\bm q,i\omega_n=0)}{i \partial q^\ra}.
	\label{eq:dm_chi}
\end{align}
Here, $(\sa,\sb,\gamma)=(x,y,z), (y,z,x),$ or $(z,x,y)$, and
$\chi_0$ is the non-interacting spin susceptibility defined as
\begin{align}
	&\chi_0^{\sa \gamma}(\bm q, i\omega_l) = -\frac{T}{V} \sum_{l,l',s_1,s_2,s_3,s_4}\sum_{\bm k, m} \sigma^\sa{s_4 s_1}\nonumber\\
	&\times G^0_{ls_1 l' s_2} (\bm k, i\omega_m) \sigma^\gamma_{s_2 s_3} G^0_{l' s_3 l s_4}(\bm k + \bm q, i\omega_m + i \omega_l),
\end{align}
where $\sigma$ is the Pauli matrix and $G^0$ is the non-interacting Green's function in the orbital basis.
Using this spin susceptibility, we can write the DM interaction as
\begin{align}
	D_\ra^\sb &= \frac{1}{V} \sum_{\bm k} D_\ra^\sb(\bm k)\\
	D_\ra^\sb(\bm k) &= 
	\lim_{\bm q \to 0} \frac{\partial}{i \partial q^\ra} \sum_{n,n'}
	\frac{f(\varepsilon_{n' \bm k+ \bm q}) - f(\varepsilon_{n \bm k})}{\varepsilon_{n' \bm k+ \bm q} - \varepsilon_{n \bm k}}\nonumber\\
	& \times \langle n\bm k| \sigma^\alpha | n' \bm k + \bm q \rangle
	\langle n'\bm k+ \bm q| \sigma^\gamma | n \bm k \rangle,
\end{align}	
where $| n \bm k \rangle$ is the eigenvector of the Kohn-Sham Hamiltonian with the eigenvalue of $\varepsilon_{n \bm k}$.
Note that this representation uses the expectation value of two Green's functions, which is similar to that using the magnetic force theorem and is in sharp contrast to the spin current approach.

\subsection{Relationship between band structures and DM interactions}
\begin{figure}
	\begin{center}
		\includegraphics[scale=0.35]{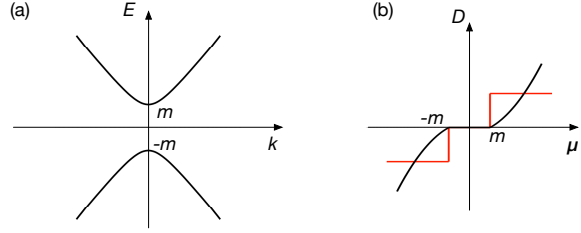}
		\caption{
			(Color online) (a) Band structure of the two-band model defined in Eq.~\eqref{eq:2x2ham} and (b) the chemical potential dependence of the DM interaction corresponding to this band structure calculated by the spin current method (black line) and spin susceptibility method (red line).
		}
	\label{fig:schematic2x2}
	\end{center}
\end{figure}

One advantage of spin current and spin susceptibility approaches is that we can discuss the relationship between band structures and the DM interaction.
In fact, we can easily calculate the contribution from the band anticrossing points, where the nontrivial spin texture arises owing to the spin-orbit couplings.
Here, let us consider the simple 2$\times$2 Hamiltonian
\begin{align}
	H = \gamma (k_x \sigma^x + k_y \sigma^y + m \sigma^z).
	\label{eq:2x2ham}
\end{align}
Then, using Eq.~\eqref{eq:dm_spincurrent} or Eq.~\eqref{eq:dm_chi},
we obtain
\begin{align}
	D_\ra^\sa = \gamma D \begin{pmatrix}
		1 & 0 \\
		0 & 1
	\end{pmatrix}
\end{align}
with
\begin{align}
	D = n_{\rm el} = \pi (\mu^2 - m^2) [ \theta(\mu - m) - \theta(-\mu - m) ]
\end{align}
from Eq.~\eqref{eq:dm_spincurrent}
and
\begin{align}
	D = \frac{\Jex^2}{16 \pi} [ \theta(\mu - m) - \theta(-\mu - m) ]
\end{align}
from Eq.~\eqref{eq:dm_chi}.
Here we assume that the number of electrons, $n_{\rm el}$, is zero at $\mu = 0$ and the row and column correspond to spatial ($\ra$) and spin ($\sa$) indices, respectively.
Schematic pictures of the energy band and the DM interactions are shown in Fig.~\ref{fig:schematic2x2}.
It is interesting to note that the DM interaction is negative for $\mu < -m$ and positive for $\mu > m$ for $\gamma > 0$,
indicating that this anticrossing point gives positive contributions to the DM interaction.
That is, when the chemical potential sweeps across the anticrossing points from below to above, the DM interaction increases.

\begin{figure}
	\begin{center}
		\includegraphics[bb=0 0 1200 600, scale=0.20]{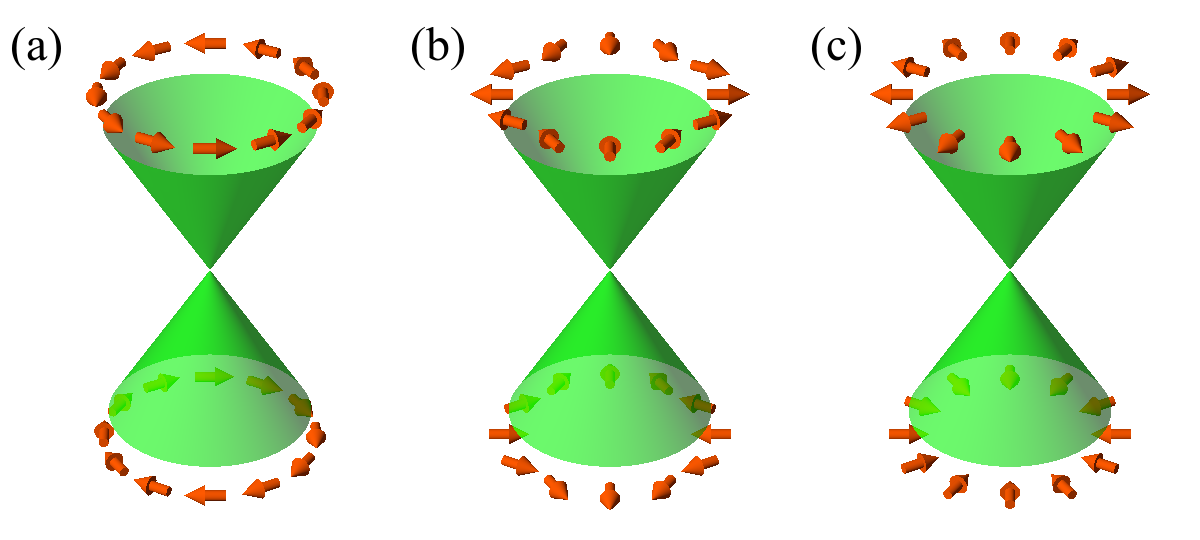}
		\caption{
			(Color online) Spin texture in momentum space for (a) Rashba, (b) Dresselhaus, and (c) Weyl-type Hamiltonians.
 Adapted with permission.\cite{kikuchi2016} Copyright 2016, American Physical Society.
		}
		\label{fig:spin_textures}
	\end{center}
\end{figure}

The relationship between the spin configurations and the symmetry of the DM coefficient is also clear in these formalisms.
Let us consider three typical spin configurations of a conduction electron in the momentum space, the Rashba (which arises in polar systems), Dresselhaus, and Weyl (in chiral systems) configurations, represented by the Hamiltonians
$H_{\rm R}=\alpha(k_x\sigma^y-k_y\sigma^x)$,
$H_{\rm D}=\beta(k_x\sigma^x-k_y\sigma^y)$, and
$H_{\rm W}=\gamma(k_x\sigma^x+k_y\sigma^y)$, respectively.
The schematic spin textures are shown in Fig.\ \ref{fig:spin_textures}.
The DM interactions in these cases (denoted by $D_{\rm R}$, $D_{\rm D}$, and $D_{\rm W}$, respectively) are
\begin{align}
	&D_{{\rm R},\ra}^\sa =  \alpha n_{\rm el} \left(\begin{array}{ccc} 0 &1&0\\ -1 & 0 & 0 \\ 0&0& 0\end{array}\right),
\;
D_{{\rm D},\ra}^\sa = \beta n_{\rm el} \left(\begin{array}{ccc} 1 &0&0\\ 0 & -1 & 0 \\ 0&0& 0\end{array}\right),
\;
\notag \\
&D_{ {\rm W},\ra}^\sa = \gamma n_{\rm el} \left(\begin{array}{ccc} 1 &0&0\\ 0 & 1 & 0 \\ 0&0& 0\end{array}\right).
\label{DR and DD and DW}
\end{align}
This result clearly relates the symmetry of crystal structures, spin-orbit couplings, and the DM interactions.
For example, DM interactions are antisymmetric in polar systems whereas diagonal DM interactions are expected in non-polar systems, as discussed by a different approach\cite{Kim13,Gungordu16}.

\section{Application to Chiral Ferromagnets}

\subsection{Electronic structure of FeGe}
FeGe is a B20-type chiral ferromagnet and is extensively studied experimentally.
A skyrmion crystal state due to the DM interaction has been observed near room temperature\cite{yu2011} and the divergence of the skyrmion size has been pointed out in \MnFeGe\ with $x=0.8$, indicating the sign change of the DM interaction\cite{shibata2013}.
Neutron scattering experiments also suggest the sign change of the DM interaction in \MnFeGe\ with $x=0.8$\cite{grigoriev2013} and \FeCoGe\ with $x=0.6$\cite{grigoriev2014}.
In MnGe, a unique three-dimensional spin structure has been observed\cite{kanazawa2011,kanazawa2012}.
The crystal symmetry of the B20 compounds is P2$_1$3, and owing to its symmetry, the DM interaction should be given as $D_\mu^\sa = D \delta_{\sa\mu}$.

\begin{figure}
	\begin{center}
		\includegraphics[scale=0.60]{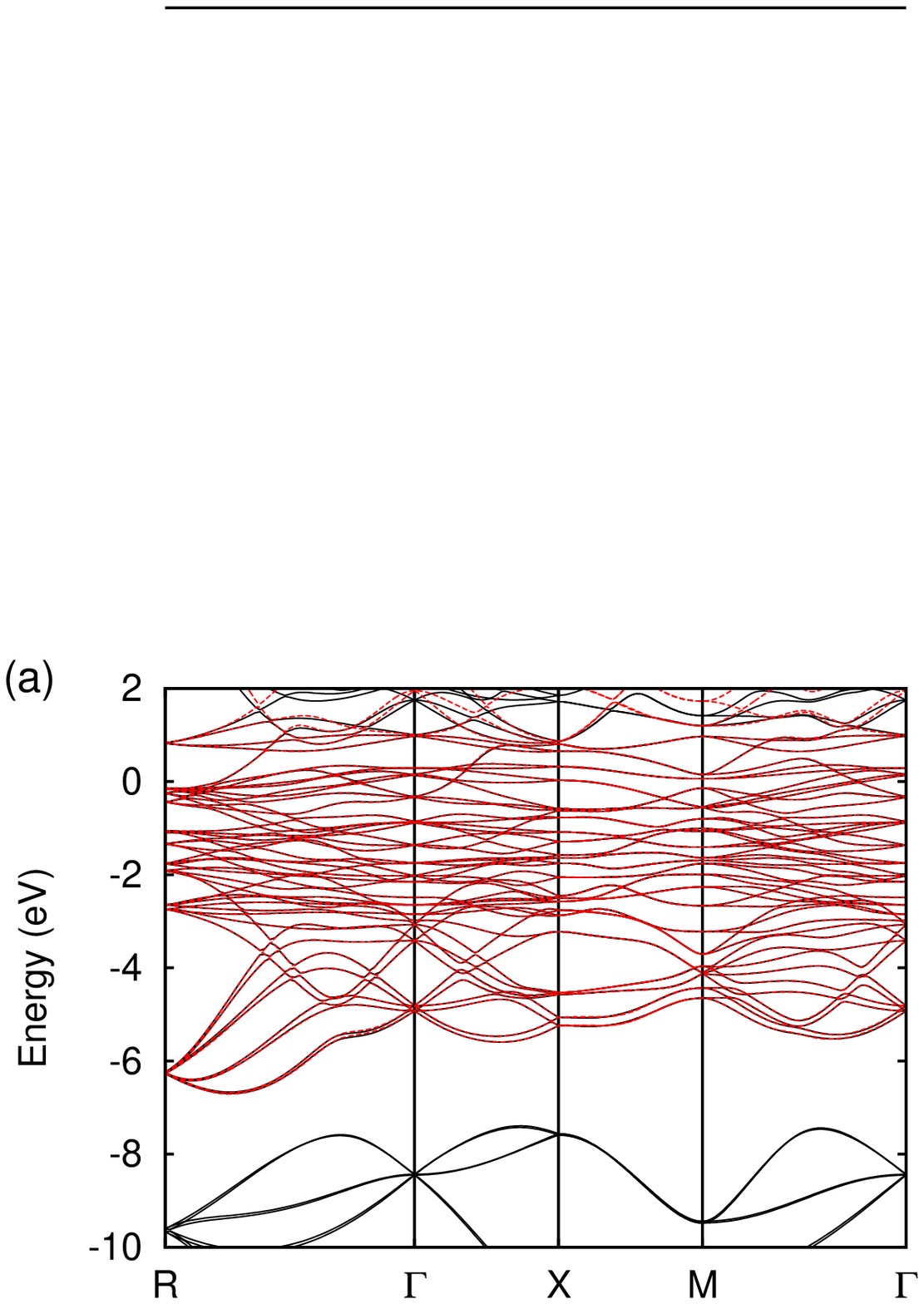}
		\includegraphics[scale=0.60]{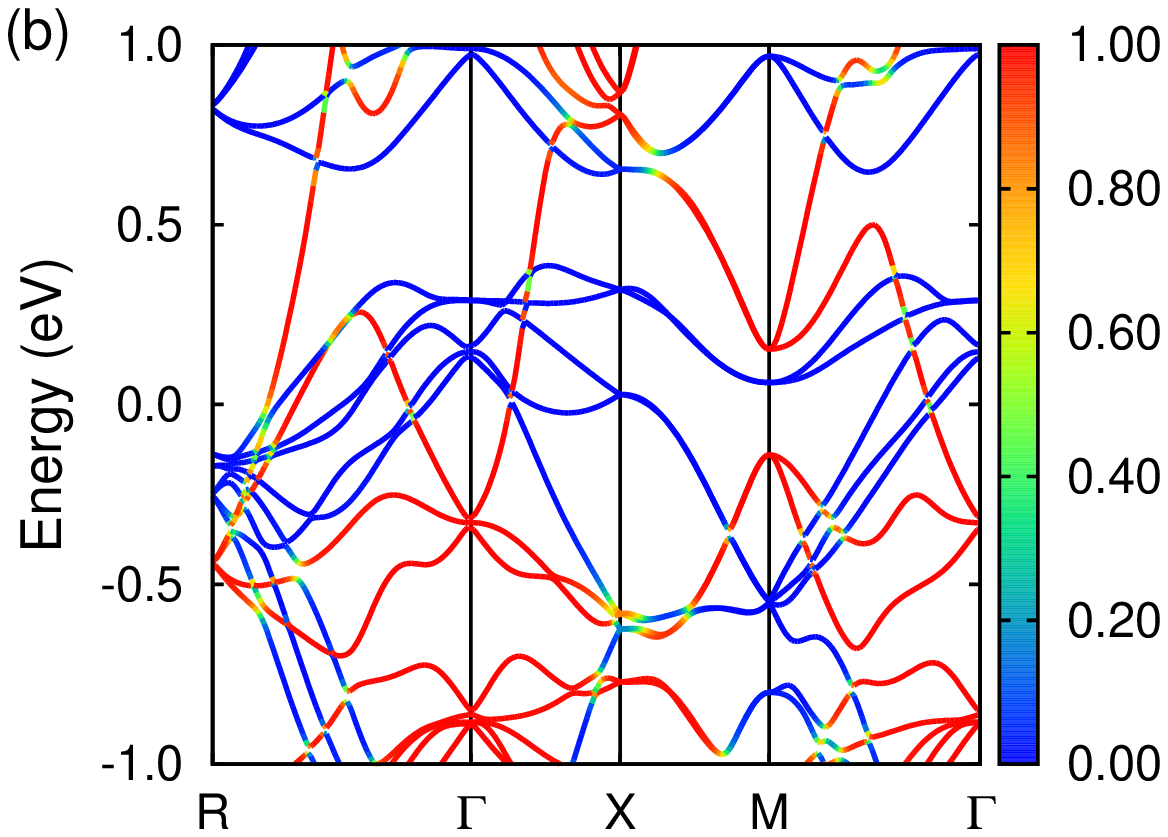}
		\caption{
			(Color online) (a) Comparison between DFT band structure (black solid lines) and tight-binding band structure (red broken lines).
			(b) Detailed band structure around the Fermi level with colors representing the weight of the up spin;
			that is, red (blue) lines correspond to up-spin (down-spin) bands.
			The Fermi level is set to zero.
			Adapted from Ref.\ \citen{koretsune2015} under the CC-BY 4.0 license.
		}
		\label{fig:band}
	\end{center}
\end{figure}

Figure \ref{fig:band}(a) shows the DFT band structure of FeGe (black solid lines).
Here, the spin-orbit couplings are included and the total ferromagnetic moment is parallel to the (001) direction ($z$ axis).
The calculated local magnetic moment is 1.18 $\mu_B$ per Fe atom, which is consistent with the results of experiments\cite{wappling1968,lundgren1968} and previous calculations\cite{yamada2003}.
The red broken lines are the Wannier-interpolated band structure with Fe 3d and Ge 4p Wannier orbitals.
According to the average energy difference between up and down spins for the Fe 3d orbitals, the exchange splitting of the 3d orbitals is estimated to be $\Delta = 1.17$ eV.
Figure \ref{fig:band}(b) shows the obtained tight-binding band structure around the Fermi level with the color representing the weight of the up spin.
Since the spin-orbit coupling of FeGe is not strong, each band is basically characterized as either an up-spin or down-spin band, and the complex spin texture emerges only around the band anticrossing region.
Hereafter, to discuss the atomic composition dependences of \MnFeGe\ and \FeCoGe, we show the results obtained using the electronic structures with self-consistent charge densities for corresponding carrier densities by fixing the atomic geometries and the lattice constant to the experimental values of FeGe\cite{lebech1989}.

\subsection{DM interaction using $E(q)$}

\begin{figure*}
	\begin{center}
		\includegraphics[scale=0.70]{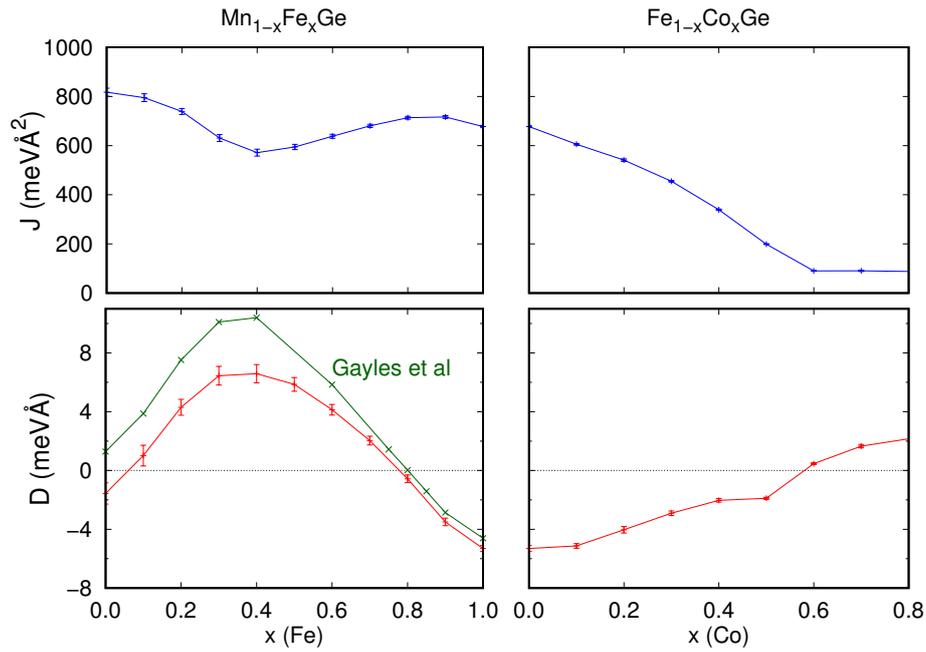}
		\caption{
			(Color online) Spin stiffness $J$ and
			DM interaction $D$ for \MnFeGe\ and \FeCoGe\ calculated using the energies of helical spin structures $E(q)$ from Ref.\ \citen{gayles2015} (green line) and Ref.\ \citen{kikuchi2016} (red lines).
			The error bars for the red lines indicate the fitting errors of $E(q) = J q^2 - D q$.
		}
		\label{fig:DM_Eq}
	\end{center}
\end{figure*}

Figure \ref{fig:DM_Eq} shows the spin stiffness and the DM interaction calculated using $E(q)$ for \MnFeGe\ and \FeCoGe$\;$ in two different studies.\cite{gayles2015,kikuchi2016} 
Although there is a slight difference in the two studies due to the treatment of the alloys, the results well reproduce the sign change of the DM interaction observed in the experiments, that is, at $x=0.8$ in \MnFeGe\ and $x=0.6$ in \FeCoGe.
In addition, a recent spin-wave spectroscopy experiment has shown that the DM interaction for FeGe is $\sim$ -3.6 meV\AA, which is also consistent with the calculation.
In contrast, around MnGe, the DM interaction is very small while the experiments show a very small skyrmion size, indicating a large DM interaction\cite{kanazawa2011}.
This discrepancy may come from the validity of the evaluation using $E(q)$ since this method assumes that the spatial spin variation is small.
In contrast with other skyrmion materials like FeGe and MnSi, MnGe is proposed to have a unique three-dimensional skyrmion lattice structure\cite{kanazawa2012}, which suggests that not only the DM interaction but also other mechanisms such as the frustration of the exchange coupling may play an important role.

The values of $J$ do not change much for \MnFeGe\ and are on the order of 1 eV\AA$^2$.
For \FeCoGe, on the other hand, the values of $J$ decrease with increasing $x$ and almost vanish for $x > 0.6$, indicating that the ferromagnetic state becomes unstable.
The wavelength of the helix, $4 \pi J/D$, for FeGe is estimated to be 160 nm, which is almost the same order as the experimental value.



\subsection{DM interaction using the spin current}

\begin{figure}
	\begin{center}
		\includegraphics[scale=0.60]{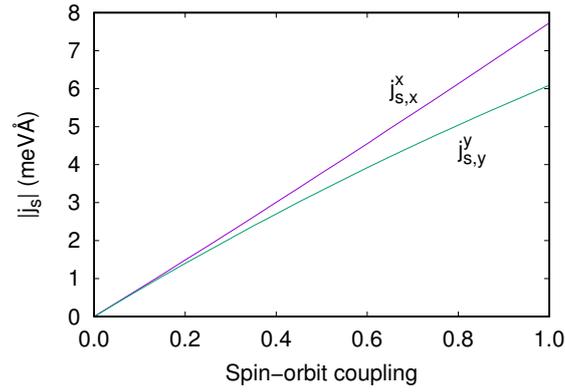}
		\caption{
			(Color online) Spin currents, $j_{{\rm s},x}^x$ and $j_{{\rm s}, y}^y$ for FeGe as a function of spin-orbit coupling strength.
		}
		\label{fig:lambda_dep}
	\end{center}
\end{figure}

\begin{figure*}
	\begin{center}
		\includegraphics[scale=0.70]{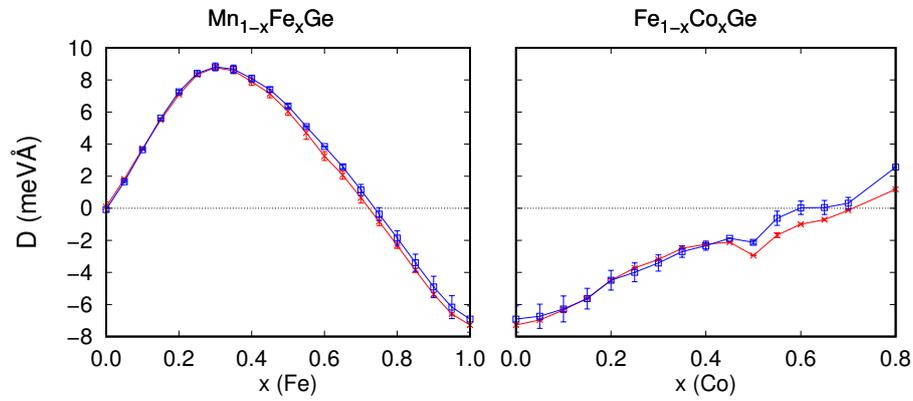}

		\caption{
			(Color online) DM interactions $D$ for \MnFeGe\ and \FeCoGe\ calculated using the spin current (blue line) and the $\lambda$-linear contribution in the spin current (red line).
			The error bars indicate the variances of $D_x^x$ and $D_y^y$.
		}
		\label{fig:DM_js}
	\end{center}
\end{figure*}

\begin{figure*}
	\begin{center}
		\includegraphics[scale=0.70]{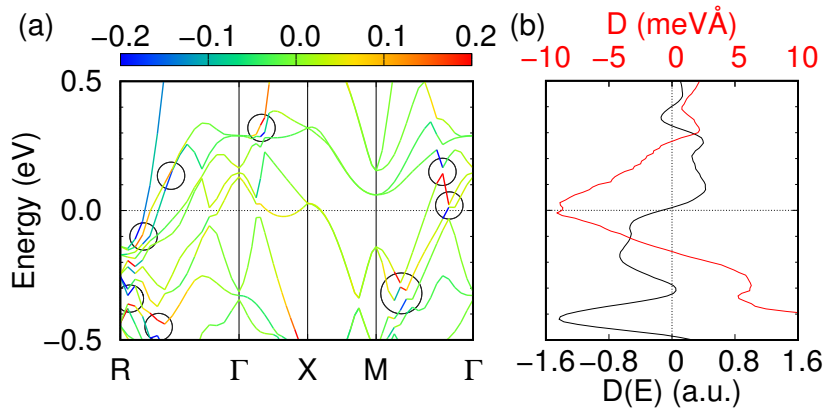}
		\caption{
			(Color online) (a) Contribution of each band to the DM interaction, $D_{n\bm k}$, with dominant band anticrossing points circled, and (b) the energy distribution of the DM interaction, $D(E)$ (black line), for FeGe. 
			The Fermi energy dependence of the DM interaction, $D\equiv\int^{E_F} D(E')f(E')dE'$, within the rigid band approximation is also shown as the red line.
			Adapted with permission.\cite{kikuchi2016} Copyright 2016, American Physical Society.
			}
		\label{fig:Dk}
	\end{center}
\end{figure*}

Next, let us discuss the DM interaction using Eq.~\eqref{J_DFT}.
Since this equation is valid within the first order of the spin-orbit coupling, it is important to check the spin-orbit-coupling-strength dependence of the spin current.
For this purpose, an electronic structure calculation without the spin-orbit coupling is performed, and a tight-binding model that reproduces this band structure is constructed.
By mixing the two tight-binding Hamiltonians with and without the spin-orbit coupling, the spin current as a function of the spin-orbit coupling strength is calculated as shown in Fig.\ \ref{fig:lambda_dep}.
As can be seen, the spin current is determined almost within the first order of the spin-orbit coupling in this system, indicating that the spin current is a good approximation for the DM interaction.
Here, the difference between $j_{{\rm s},x}^x$ and $j_{{\rm s},y}^y$ originates from the magnetic moment, which breaks the symmetries that connect the $x$- and $y$-axes.
In fact, if we consider the magnetic moment along the (111) direction, the two perpendicular spin currents have the same values due to the C$_3$ symmetry.
Note that for large spin-orbit-coupling systems such as a Co/Pt bilayer, the higher-order contributions cannot be neglected\cite{freimuth2017}.

Figure \ref{fig:DM_js} shows the DM interaction estimated directly using the spin current and by extracting the first-order contribution from the spin current.
Since the idea of the spin current approach is the same as the method using $E(q)$, these results give almost the same result as shown in Fig.\ \ref{fig:DM_Eq}.
The agreement between the first-order contribution and the original spin current supports the validity of this estimation.
Interestingly, the difference between $j_{{\rm s},x}^x$ and $j_{{\rm s},y}^y$ becomes smaller for the first-order contribution.
The higher-order contribution of the spin-orbit coupling and the possible anisotropy of the DM interaction are important future issues.

In the spin current approach, it is possible to discuss the relationship between the band structure and the DM interaction and, as a result, the chemical potential dependence of the DM interaction.
In fact, we can rewrite Eq.~\eqref{J_DFT} as
\begin{align}
	D = \sum_{n \bm k} D_{n\bm k} f(\epsilon_{n\bm k}) = \int D(E) f(E) dE,
\end{align}
where $n$, $f(E)$, $D_{n\bm k}$, and $D(E)$ are the band index, the Fermi distribution function, the contribution of each band to the DM interaction, and the density of the DM interaction, respectively.
Figure \ref{fig:Dk}(a) shows the band structure of FeGe with the color representing $D_{n\bm k}$.
As discussed in Sect. 2, we can see that the DM interaction comes from the restricted region of the band structure where the complex spin texture arises due to band anticrossing.
The density of the DM interaction, $D(E)$, shown in Fig.\ \ref{fig:Dk}(b), also gives useful information for discussing the carrier density dependence of the DM interaction. 
That is, in this case, $D(E) < 0$ for $E<0$ and $D(E) > 0$ for $E>0$ indicate the dip structure around FeGe ($E=0$) and the resulting two sign changes in \MnFeGe\ and \FeCoGe.

\subsection{DM interaction using the spin susceptibility}

\begin{figure*}
	\begin{center}
		\includegraphics[scale=0.70]{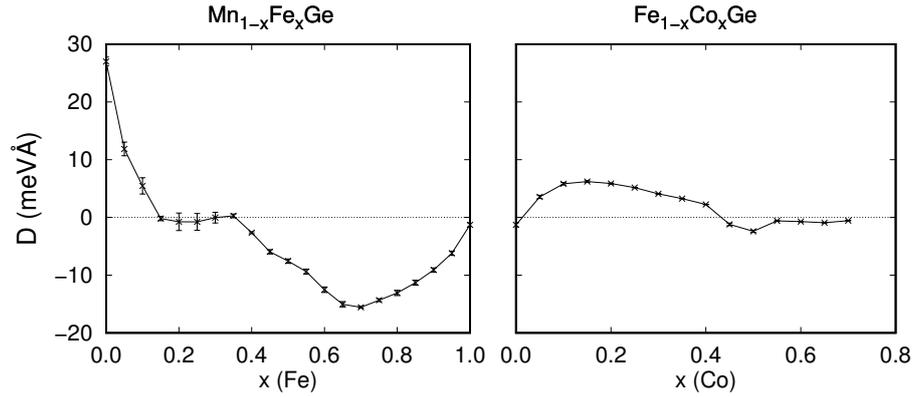}
		\caption{
			DM interactions for \MnFeGe\ and \FeCoGe\ calculated using the off-diagonal spin susceptibility.
			The error bars indicate the variances of $D_x^x$ and $D_y^y$.
		}
		\label{fig:DM_chiq}
	\end{center}
\end{figure*}

The DM interaction using Eq.~\eqref{eq:dm_chi} is shown in Fig.\ \ref{fig:DM_chiq}.
Here, the exchange coupling $\Jex$ is estimated by the exchange splitting of the $d$ orbitals, that is, the average energy difference of $3d$ Wannier orbitals for up and down spins.
In this result, the two sign-change points ($x\sim0.4$ for \MnFeGe\ and $x\sim0.02$ for \FeCoGe) are shifted to some extent from the positions in the experiments.
This may be because the perturbation expansion with respect to $\Jex$ is not a good approximation in this material.
In fact, $\Jex$ is not so small and the effect of exchange splitting on the electronic structure cannot be explained only by $\Jex {\bm n} \cdot \sigmav$.
On the other hand, the DM interaction for \MnFeGe\ with small $x$ is larger than the previous two results and is more consistent with the experimental result.
This suggests that the pertubation expansion with respect to $\Jex$ reasonably works even for large $\Jex$ systems and can give a better estimation once the derivative expansion with respect to the magnetic structure does not work.
In this way, this method can play a complementary role in the evaluation of the DM interaction.

\section{Conclusion}
In this paper, we have reviewed three approaches to evaluate the DM interaction.
The first one is to evaluate the energy of the helical spin structure, $E(\bm q)$, directly from first principles and extract the DM interaction.
The method is very powerful while the obtained information is limited.
The second one is the perturbation expansion with respect to the spin gauge field, which represents the spin twisting.
The idea is to extract the DM interaction term by twisting the spin structure, which is almost the same as the first approach, while this method only needs the electronic structure calculation for the uniform magnetic state.
This method gives a clear picture that the DM interaction can be expressed in terms of the spin current at the equilibrium within the first order of the spin-orbit couplings.
Furthermore, the relationship to the band structure and the effect of carrier doping can be easily discussed.
The third one is the perturbation expansion with respect to the exchange coupling, which can be understood as the extension of the RKKY mechanism.
This method also clarifies the relationship between the band structure and the DM interaction, which is roughly consistent with the spin current approach.

By applying these methods to chiral ferromagnets, it has been shown that the first two approaches give almost the same results, which agree well with the experiments, while the third one gives a slightly different result.
Since the starting points of the first two approaches and the third approach are completely different, these approaches will play complementary roles in the evaluation of the DM interaction.
The application of these methods to various systems and clarifying their validity are important future issues for designing materials utilizing this unique antisymmetric interaction.


\acknowledgement
The authors thank H. Fukuyama, N. Kanazawa, H. Kawaguchi, W. Koshibae, H. Kohno, T. Momoi, D. Morikawa, N. Nagaosa, M. Ogata, S. Seki, K. Shibata, Y. Suzuki, G. Tatara, and Y. Tokura for valuable discussions.
In particular, the authors thank G. Tatara for the collaboration in Ref.\ \citen{kikuchi2016} and T. Ko. and R. A. thank N. Nagaosa for the collaboration in Ref. \citen{koretsune2015}.
This work was supported by JSPS KAKENHI Grant Numbers JP25400344 and JP26103006 and JST PRESTO Grant Number JPMJPR15N5.  T. Ki. is a Yukawa Research Fellow supported by the Yukawa Memorial Foundation.    

\bibliographystyle{jpsj}
\bibliography{81712}

\end{document}